\documentclass[twocolumn,times]{aastex63}

\newcommand\cts{counts~s$^{-1}$}
\newcommand\ctscm{counts~s$^{-1}$~cm$^{-2}$}

\def\CC{C\raise.22ex\hbox{{\footnotesize +}}\raise.22ex\hbox{\footnotesize +}}

\shorttitle{Modeling the in-orbit background of PolarLight}
\shortauthors{Huang et al.}

\begin{document}

\title{Modeling the in-orbit background of PolarLight}

\author{Jiahui Huang}
\affiliation{Department of Engineering Physics and Center for Astrophysics, Tsinghua University, Beijing 100084, China}

\correspondingauthor{Hua Feng}
\email{hfeng@tsinghua.edu.cn}

\author[0000-0001-7584-6236]{Hua Feng}
\affiliation{Department of Astronomy, Tsinghua University, Beijing 100084, China}
\affiliation{Department of Engineering Physics and Center for Astrophysics, Tsinghua University, Beijing 100084, China}

\author{Hong Li}
\affiliation{Department of Engineering Physics and Center for Astrophysics, Tsinghua University, Beijing 100084, China}

\author{Xiangyun Long}
\affiliation{Department of Engineering Physics and Center for Astrophysics, Tsinghua University, Beijing 100084, China}

\author{Dongxin Yang}
\affiliation{Department of Engineering Physics and Center for Astrophysics, Tsinghua University, Beijing 100084, China}

\author{Weihe Zeng}
\affiliation{Department of Engineering Physics and Center for Astrophysics, Tsinghua University, Beijing 100084, China}

\author{Qiong Wu}
\affiliation{Department of Engineering Physics and Center for Astrophysics, Tsinghua University, Beijing 100084, China}

\author{Weichun Jiang}
\affiliation{Key Laboratory for Particle Astrophysics, Institute of High Energy Physics, Chinese Academy of Sciences, Beijing 100049, China}

\author{Massimo Minuti}
\affiliation{INFN-Pisa, Largo B. Pontecorvo 3, 56127 Pisa, Italy}

\author{Enrico Costa}
\affiliation{IAPS/INAF, Via Fosso del Cavaliere 100, 00133 Rome, Italy}

\author{Fabio Muleri}
\affiliation{IAPS/INAF, Via Fosso del Cavaliere 100, 00133 Rome, Italy}

\author{Saverio Citraro}
\affiliation{INFN-Pisa, Largo B. Pontecorvo 3, 56127 Pisa, Italy}

\author{Hikmat Nasimi}
\affiliation{INFN-Pisa, Largo B. Pontecorvo 3, 56127 Pisa, Italy}

\author{Jiandong Yu}
\affiliation{School of Electronic and Information Engineering,  Ningbo University of Technology, Ningbo, Zhejiang 315211, China}

\author{Ge Jin}
\affiliation{North Night Vision Technology Co., Ltd., Nanjing 211106, China}

\author{Zhi Zeng}
\affiliation{Department of Engineering Physics and Center for Astrophysics, Tsinghua University, Beijing 100084, China}

\author{Ming Zeng}
\affiliation{Department of Engineering Physics and Center for Astrophysics, Tsinghua University, Beijing 100084, China}

\author{Peng An}
\affiliation{School of Electronic and Information Engineering,  Ningbo University of Technology, Ningbo, Zhejiang 315211, China}

\author{Luca Baldini}
\affiliation{INFN-Pisa, Largo B. Pontecorvo 3, 56127 Pisa, Italy}

\author{Ronaldo Bellazzini}
\affiliation{INFN-Pisa, Largo B. Pontecorvo 3, 56127 Pisa, Italy}

\author{Alessandro Brez}
\affiliation{INFN-Pisa, Largo B. Pontecorvo 3, 56127 Pisa, Italy}

\author{Luca Latronico}
\affiliation{INFN, Sezione di Torino, Via Pietro Giuria 1, I-10125 Torino, Italy}

\author{Carmelo Sgr\`{o}}
\affiliation{INFN-Pisa, Largo B. Pontecorvo 3, 56127 Pisa, Italy}

\author{Gloria Spandre}
\affiliation{INFN-Pisa, Largo B. Pontecorvo 3, 56127 Pisa, Italy}

\author{Michele Pinchera}
\affiliation{INFN-Pisa, Largo B. Pontecorvo 3, 56127 Pisa, Italy}

\author{Paolo Soffitta}
\affiliation{IAPS/INAF, Via Fosso del Cavaliere 100, 00133 Rome, Italy}

\begin{abstract}

PolarLight is a gas pixel X-ray polarimeter mounted on a CubeSat, which was launched into a Sun-synchronous orbit in October 2018. We build a mass model of the whole CubeSat with the Geant4 toolkit to simulate the background induced by the cosmic X-ray background (CXB) and high energy charged particles in the orbit.  The simulated energy spectra and morphologies of event images both suggest that the measured background with PolarLight is dominated by high energy electrons, with a minor contribution from protons and the CXB. The simulation reveals that, in the energy range of 2--8 keV, there are roughly 28\% of the background events are caused by energy deposit from a secondary electron with an energy of a few keV, in a physical process identical to the detection of X-rays. Thus, this fraction of background cannot be discriminated from X-ray events.  The background distribution is uneven on the detector plane, with an enhancement near the edges. The edge effect is because high energy electrons tend to produce long tracks, which are discarded by the readout electronics unless they have partial energy deposits near the edges. The internal background rate is expected to be around $6 \times 10^{-3}$~\ctscm\ in 2--8 keV if an effective particle discrimination algorithm can be applied. This indicates that the internal background should be negligible for future focusing X-ray polarimeters with a focal size in the order of mm.  

\end{abstract}


\section{Introduction}

PolarLight is a dedicated soft X-ray polarimeter onboard a CubeSat, which was launched into a Sun-synchronous orbit on 29 October 2018 \citep{Feng2019,Feng2020}. The X-rays are measured by the gas pixel detector (GPD), a 2D position-sensitive gas proportional counter that enables high-sensitivity X-ray polarimetry on the basis of the photoelectric effect \citep{Costa2001,Bellazzini2013}.  The GPD measures the 2D trajectory of photoelectrons following the absorption of X-rays in the gas chamber, and the X-ray polarization can be inferred via statistical analysis of the emission angle of photoelectrons on the detector plane. The purpose of PolarLight is to demonstrate the new technique in space, perform scientific observations, and obtain a better understanding of the in-orbit background.  The mission profile and ground calibrations are elaborated in \citet{Feng2019}. Operation and status of the instrument in the orbit can be found in \citet{Li2021}. Scientific results with observations of the Crab nebula are reported in \citet{Feng2020a}.  This paper focuses on the modeling of the in-orbit background. 

Understanding the in-orbit background is of great importance for astronomical high energy instruments, as the sensitivity of instruments is immediately determined by the background.  PolarLight is a non-imaging detector,  whose background estimate and subtraction is always challenging \citep{Jahoda2006,Fitzpatrick2012}. Mass modeling is widely used to interpret or predict the background of space-borne high energy instruments \citep{Dean2003,Shaw2003,Ferguson2003,Xie2015,Antia2017,Ripa2019}. PolarLight is the first of its kind for soft X-ray polarimetry in space. Thus, study of its background would help optimize the scientific returns of future missions like the Imaging X-ray Polarimetry Explorer \citep[IXPE;][]{Weisskopf2016} and the enhanced X-ray Timing and Polarimetry \citep[eXTP;][]{Zhang2019}.  

PolarLight has been scheduled to measure the in-orbit background while the science targets are occulted by the Earth. In that case, the detector points roughly at the anti-Sun direction and rocks with a half angle of about 30$^\circ$. Therefore, a random sky region or the Earth's atmosphere is observed.  The field of view (FOV) of PolarLight is narrow, with a full width of 2.3$^\circ$ at half maximum. Due to the small area, PolarLight is only sensitive to sources as bright as the Crab nebula. We will show later that the in-orbit background is mainly caused by charged particles.  Thus, observing at a random direction is equivalent of observing the pure background.  

In this paper, we try to interpret the in-orbit background of PolarLight using the mass modeling technique.  The sensitive range for energy deposits by X-rays and charged particles is 1--10 keV, while the energy band sensitive to X-ray polarimetry is 2--8 keV. Throughout the paper, we study and quote the background properties in the full band of 1--10 keV, but use the 2--8 keV band if we want to evaluate the effect of background in polarization measurement.  As of March 2020, background observations with a total exposure of 230~ks have been accumulated, with a mean count rate of about 0.19~\cts\ in the full band or 0.07~\cts\ in 2--8~keV. The paper is organized as follows. The mass model and simulations are described in \S~\ref{sec:model}. The simulation results and comparisons with the measurements are shown in \S~\ref{sec:result} and discussed in \S~\ref{sec:discussion}.

\section{Model and simulation}
\label{sec:model}

The Geant4 package\footnote{\url{https://geant4.web.cern.ch}} version 4.10.03 \citep{Agostinelli2003} is utilized for mass modeling and simulation.  It is a Monte-Carlo toolkit allowing for particle tracking in user defined geometries.  The Livermore models in Geant4 are adopted for low energy electromagnetic processes.  X-ray polarization and Auger electrons are enabled.  

\begin{figure}
\centering
\includegraphics[width=0.8\columnwidth]{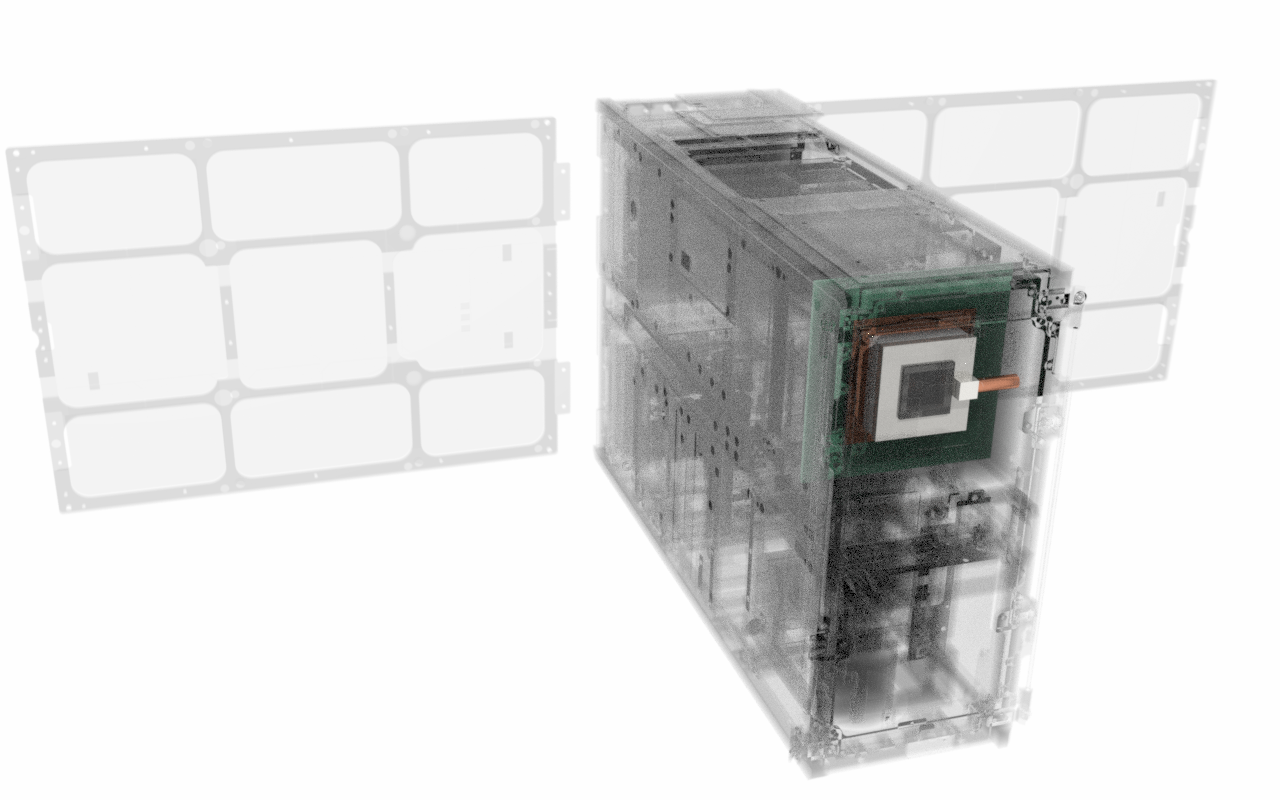}
\caption{The mass model of the CubeSat with the GPD highlighted in color. 
\label{fig:model}}
\end{figure}

The mass model of the GPD, which is the most sensitive volume in the detection, is defined using \CC\ codes in Geant4 directly, to reflect each component as precisely as possible.  The mass model of the rest of the CubeSat is created using the Geometry Description Markup Language (GDML) files imported from a mechanical model of the CubeSat, which may have some degree of simplification and have ignored small components like screws, but is sufficiently accurate in view of mass distribution.  The mass model of the CubeSat including the GPD in Geant4 is displayed in Figure~\ref{fig:model}. 

\begin{figure}
\centering   
\includegraphics[width=\columnwidth]{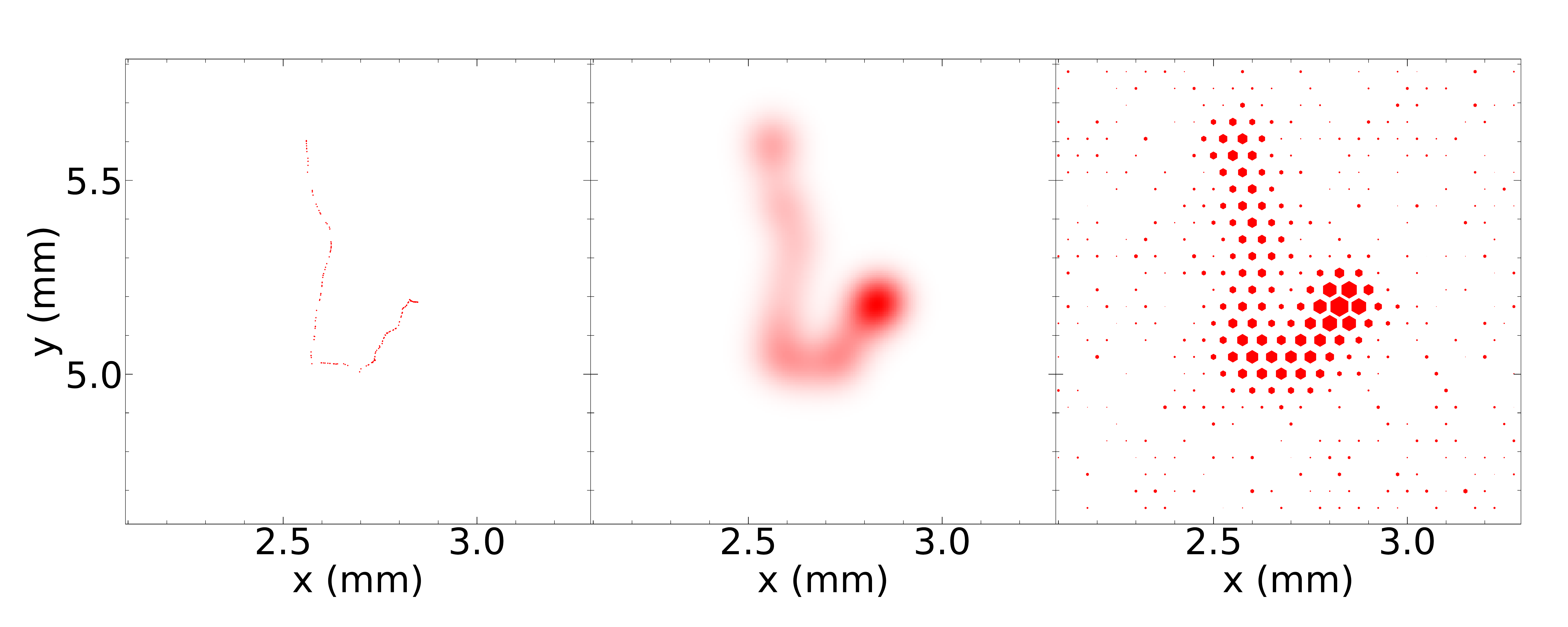}   
\caption{Illustration of how a track image is simulated. (a) Energy deposits in the gas chamber projected on the detector plane following the absorption of an 8~keV X-ray. (b) Energy distribution after Gaussian convolutions to account for transversal diffusion of charges during drift and collection, to mimic charge transportation in the gas chamber. (c) Energies integrated on hexagonal pixels with random fluctuation and readout noise.
\label{fig:sim}}
\end{figure}

Energies deposited in the sensitive volume of the detector, i.e., the drift region of the gas chamber, are recorded in Geant4 simulations.  To account for transversal thermal diffusion of electrons along drift, a Gaussian dilution is applied to charges at each deposit point with a kernel size scaled with the square root of the distance to the gas electron multiplier (GEM) foil.  After the electrons permeate the GEM foil and get multiplied, there is an additional Gaussian diffusion independent of the position.  Then, to simulate the charge collection on the hexagonal pixels of the application-specific integrated circuit (ASIC) chip, charges are integrated within pixels.  Fluctuation of the signal and the electronic noise are added on each pixel.  The parameters such as the Gaussian kernel size and noise level are adopted from measurements in the laboratory and space. The above processes are shown in Figure~\ref{fig:sim} to illustrate how a track is simulated. Such tracks are the science data products in the measurements.  We note that there is no morphology-based filtering algorithm embed in the payload. 

The background of PolarLight may result from the cosmic X-ray background (CXB), albedo X-rays from the Earth's atmosphere, and high energy charged particles in the orbit.  The CXB spectrum measured with HEAO-1 \citep{Gruber1999} is adopted.  The direct detection of CXB through the aperture of the collimator is calculated to be negligible due to the small FOV. The simulation reveals that CXB photons with energies in the range of $\sim$30--300 keV may effectively ``leak'' into the detector with energy deposits in the band of our interest. The total flux of albedo X-rays \citep{Dean2003} from the Earth's atmosphere is orders of magnitude lower than that of CXB in the above energy range, and can thus be ignored.  The shadowing of CXB by the Earth is also considered. 

The energy spectra of high energy protons and electrons\footnote{In this paper, high energy electrons in the space actually refer to both electrons and positrons.} consist of two components respectively, the primary and secondary components; particles in the latter are generated by those in the former interacting with the atmosphere, and dominate the spectrum at energies below $\sim$GeV.  We adopt the spectral models constructed based on measurements with the Alpha Magnetic Spectrometer (AMS) in \citet{Mizuno2004}, where the primary spectrum is described with a low-energy cutoff power-law and the secondary spectrum is a broken power-law.  The cutoff energy of the primary component and the spectrum of the secondary component are modulated by the magnetic rigidity and is a function of the magnetic latitude ($\theta_{\rm m}$).  The proton and electron spectra are available in several magnetic latitude bins from 0 to 0.6~rad.  The energy spectra cover an energy range from 10~MeV to 100~GeV.  The solar activity only affects the primary component at energies below $\sim$GeV, where the secondary component dominates the spectrum. As a consequence, the \citet{Mizuno2004} model predicts almost no dependence of the charged particle spectra on solar activity. 

For protons and electrons, we execute Geant4 simulations in 5 magnetic latitude ranges, $\theta_{\rm m} =$ 0--0.2, 0.2--0.3, 0.3--0.4, 0.4--0.5, and 0.5--0.6~rad, respectively.  In each magnetic latitude bin, the simulation runs with an equivalent acquisition time of 1~hr. We find that the shape of the background spectrum in the energy band of 1--10 keV, either simulated or measured, does not depend on the magnetic latitude.  Therefore, the simulated energy spectra at different magnetic latitudes can be co-added. 

\section{Results}
\label{sec:result}

\subsection{Decomposition of the background}

\begin{figure}
\centering   
\includegraphics[width=\columnwidth]{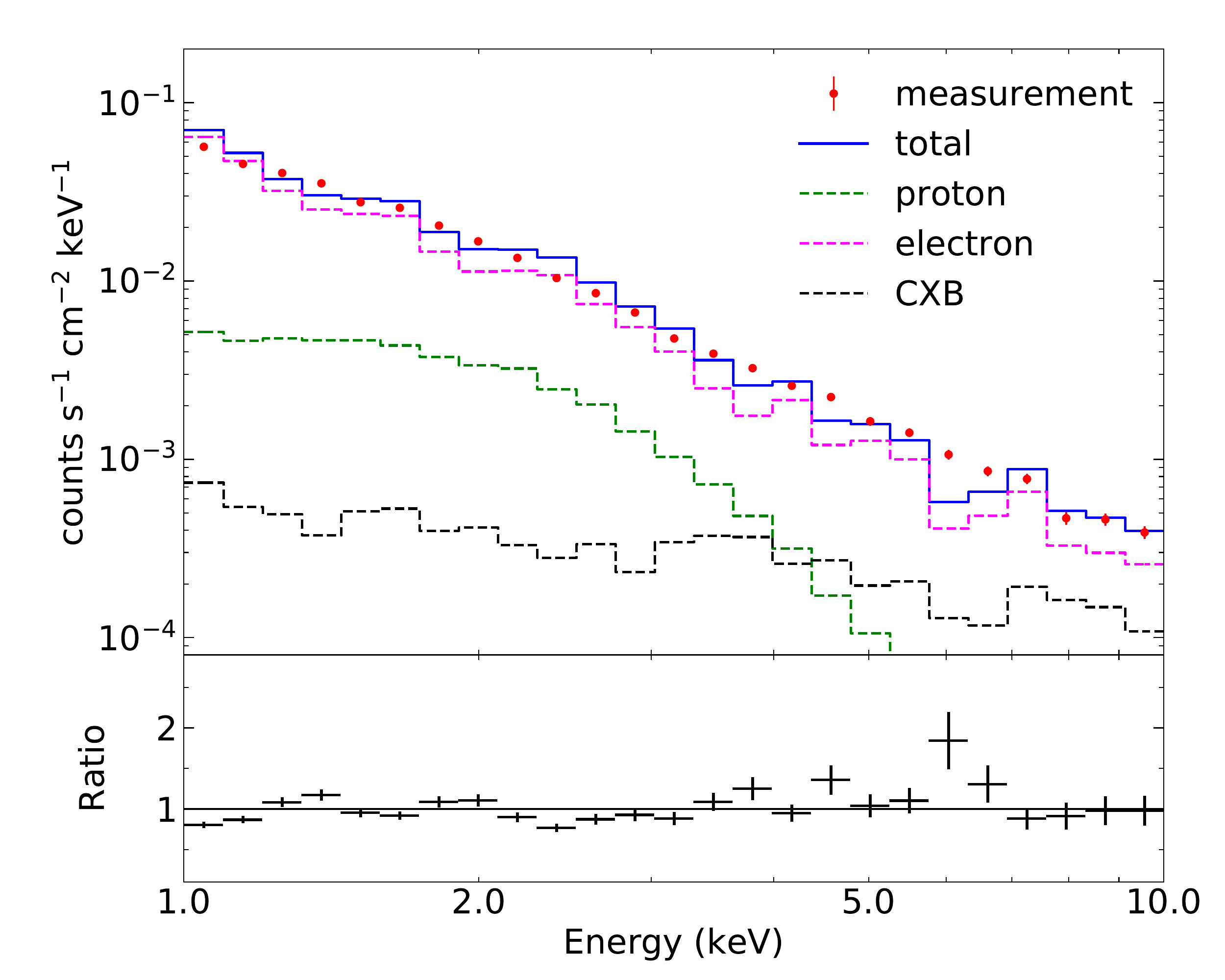}   
\caption{Measured and simulated background spectra.  The red points with errors are the measurement. The solid blue curve is the simulated background, which is decomposed into three components, respectively due to the CXB, protons, and electrons. The bottom panel is the measurement-to-simulation ratio. 
\label{fig:cmp_spec}}
\end{figure}

The simulated spectra are plotted in Figure~\ref{fig:cmp_spec} against the measured energy spectrum.  The flux of CXB is constant with time and less uncertain, while the flux of high energy charged particles may vary as a function of time and location.  We add the three simulated components to fit the measurement. The CXB component is directly adopted from simulation, while the normalizations of the proton- and electron-induced spectra are allowed to vary, to account for possible changes in their relative fluxes. It is obvious that the electron-induced background is the dominant component. It occupies 81\% of the background counts in 1-10 keV and 76\% in 2-8 keV. The fractions are 15\% and 17\% for the proton-induced background, and 4\% and 7\% for the leaked X-ray background, respectively in the two energy bands. 

\begin{figure}
\centering
\includegraphics[width=\columnwidth]{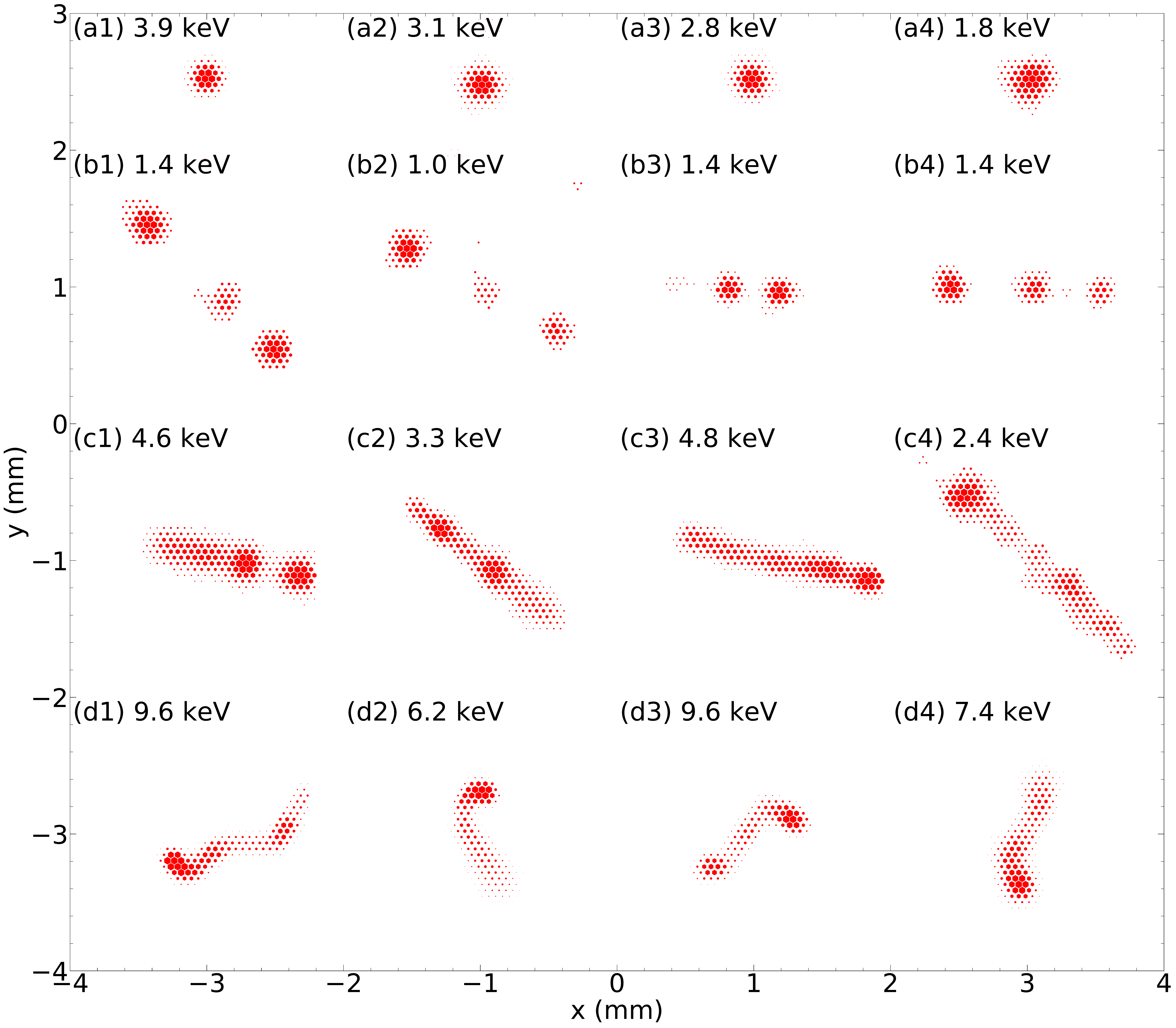}
\caption{Typical track images from simulations and measurements in each morphology category. Images from column (1) to (3) are selected, respectively, from simulations with protons, electrons, and the CXB, and images in column (4) are from in-orbit measurements of the background. The four rows are respectively for the four morphology categories: (a) point-like with a single island,  (b) point-like with multiple islands, (c) elongated and straight, and (d) elongated and curved.  The energy of each event is indicated.
\label{fig:track}}
\end{figure}

The second approach to perform the test is to compare the morphology of track images between measured and simulated data. Inspection of the track images with trained eyes, we classify the images (after noise cut) into four categories according to their morphologies, which are (a) point-like with a single island,  (b) point-like with multiple islands, (c) elongated and straight, and (d) elongated and curved.  Typical examples for events in the four categories are displayed in Figure~\ref{fig:track}.  Islands in images of type (b) are mostly aligned along a straight line.  

The mechanisms that produce tracks in the four categories can be found from simulations. The energies of incident charged particles range from 10~MeV to 100~GeV.  Both high energy protons and electrons can produce the four morphology types shown in Figure~\ref{fig:track}, but with different probabilities.  Protons with energies of $\sim$GeV or above tend to produce discrete energy deposits such as events of type (a) and (b), while those of relatively low energies may cause continuous energy loss and produce event images of type (c). Electrons typically yield continuous energy loss as events in type (c). However, in some cases, events in type (c) may appear discrete like those in type (b) or even type (a), because some parts have a low surface density and is not shown after the noise cut.  Events with images of type (c) could also be resulted from secondary electrons with energies of tens to hundreds of keV. Images of type (d) are due to secondary electrons with energies of several keV.  The leaked CXB background is mainly due to secondary electrons of tens to hundreds of keV, so that events of type (b) have a low fraction.  On the other side, X-rays through the aperture in the energy band of interest always produce images belong to type (a) and (d).   Therefore, this part of background cannot be discriminated from X-ray events. 

We randomly select 1000 events from each pool of the data (the observation and simulations with different types of particles) and calculate the fraction of events in each category in the two energy ranges (see Table~\ref{tab:fraction}). The results suggest that the simulation with the CXB or protons alone does not match the data, while the electron-induced background, as well as a combination of the three components with the best-fit fractions, can reasonably fit the observation.  In the energy range of 1--10 keV, the combined simulation results provide the best fit. In 2--8 keV, the combined simulation seems to over-estimate the type (a) and under-estimate the type (c) and (d), but we note that images of type (a) could be confused with types (c) and (d) if the track has a relatively short length.  Compared with the CXB or protons, the electron or combined results provide a much better fit. To conclude, the image morphology further confirms the decomposition of the background sources, suggesting that the in-orbit background is dominantly electron-induced.  In the following work, simulations in combination with the CXB and charged particles with the best-fit fractions are used to compare with the measurement.

\begin{deluxetable}{lrrrr}[tb]
\tablecaption{Fraction of events in different morphologies.
\label{tab:fraction}}
\tablewidth{\columnwidth}
\tablecolumns{5}
\tablehead{
\colhead{morphology category} & \colhead{(a)} & \colhead{(b)} & \colhead{(c)} & \colhead{(d)}
}
\startdata
\noalign{\smallskip}
\multicolumn{5}{c}{1--10 keV} \\
\noalign{\smallskip}\hline\noalign{\smallskip}
measurement     & 24.2\% & 17.8\% & 51.6\% & 6.4\% \\
simulation (protons)   & 53.0\% & 35.0\% & 10.3\% & 1.7\% \\
simulation (electrons) & 18.2\% & 12.8\% & 65.8\% & 3.2\% \\
simulation (CXB)        & 35.1\% & 2.7\% & 26.7\% & 35.5\% \\
simulation (combined)        & 24.0\% & 15.6\% & 56.1\% & 4.3\% \\
\noalign{\smallskip}\hline\noalign{\smallskip}
\multicolumn{5}{c}{2--8 keV} \\
\noalign{\smallskip}\hline\noalign{\smallskip}
measurement     & 16.6\% & 11.1\% & 60.8\% & 11.5\%\\
simulation (proton)   & 45.0\% & 40.7\% & 11.8\% & 2.5\%\\
simulation (electron) & 17.1\% & 9.9\% & 66.2\% & 6.8\%\\
simulation (CXB)        & 35.1\% & 1.9\% & 24.7\% & 38.3\% \\
simulation (combined)        & 23.2\% & 14.6\% & 53.8\% & 8.4\% \\
\enddata
\tablecomments{Morphology categories: (a) point-like with a single island,  (b) point-like with multiple islands, (c) elongated and straight, and (d) elongated and curved. }
\end{deluxetable}

\subsection{Orbital variation of the background}

\begin{figure}
\centering
\includegraphics[width=\columnwidth]{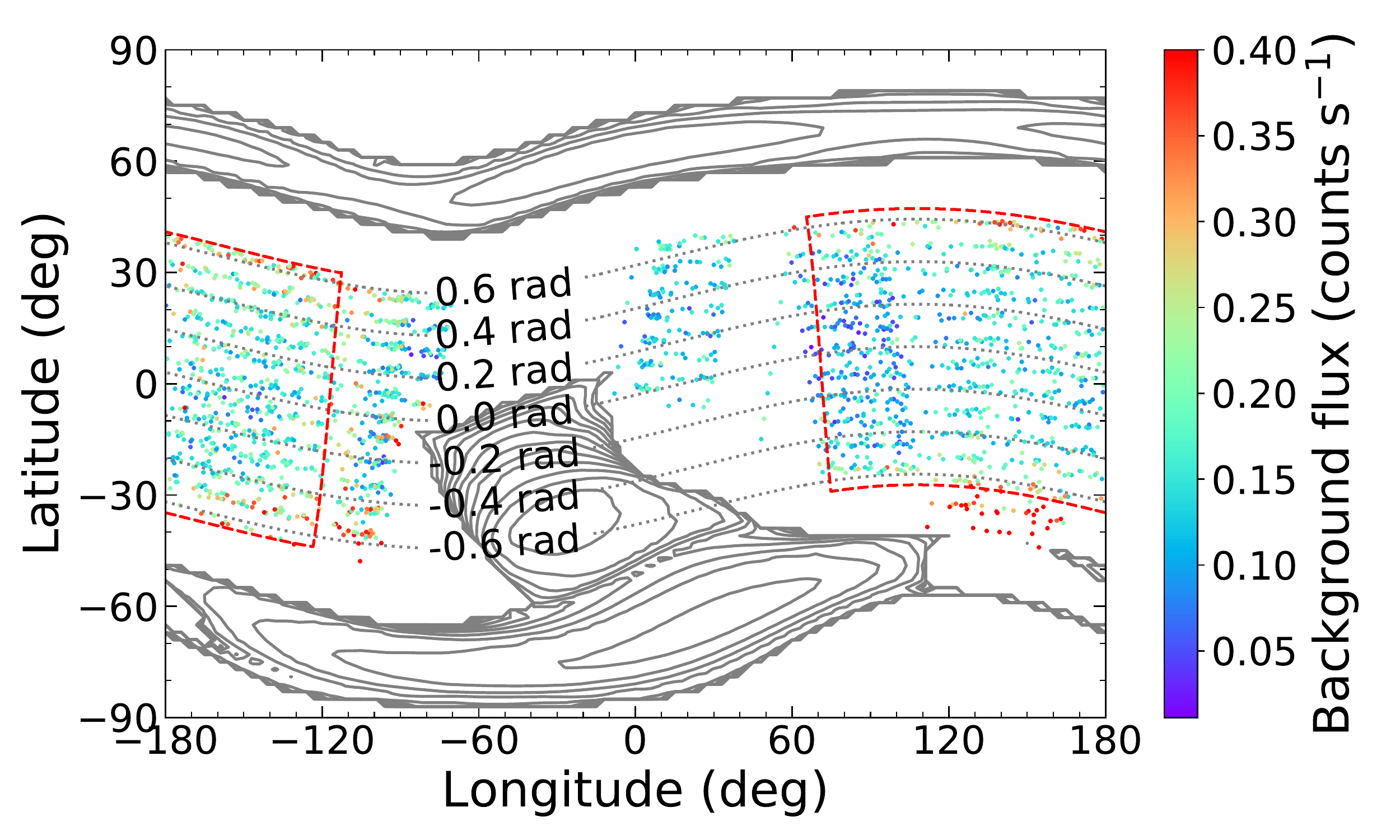}
\caption{1--10 keV background count rate (points) on top of the flux map (contours) of trapped high energy ($>100$~keV) electrons in the orbital plane of PolarLight.  Points indicate measured background count rates in every 100 seconds. The dashed lines encircle regions away from locations with trapped electrons.
\label{fig:map}}
\end{figure}

\begin{figure}
\centering
\includegraphics[width=0.8\columnwidth]{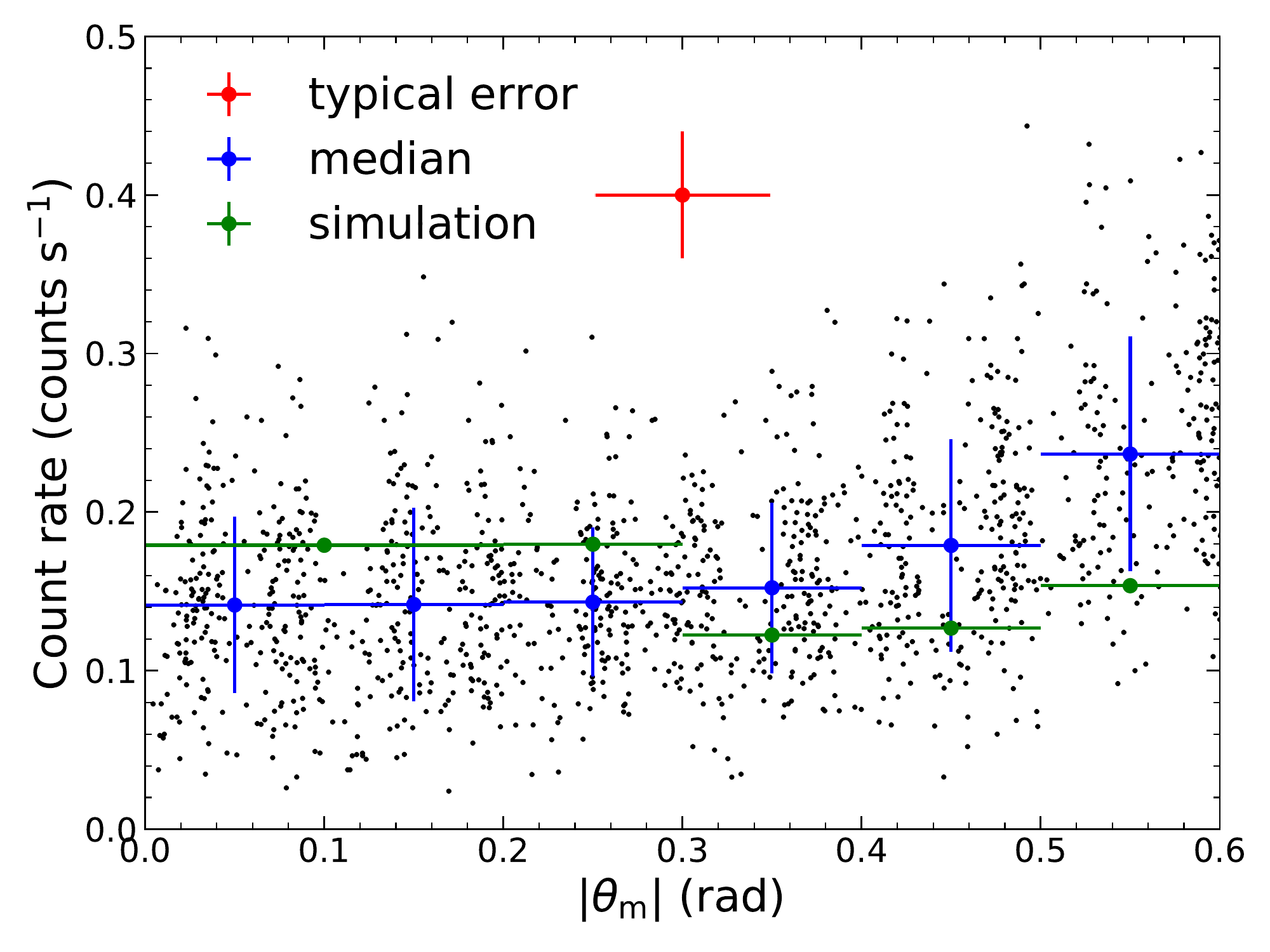}
\caption{Count rate of measured background (black dots) in 1--10 keV as a function of the magnetic latitude, from measurements in regions encircled by dashed lines in Figure~\ref{fig:map}.  The red symbol with bars indicates the typical error and span of magnetic latitude of each measurement.  The blue symbols are the median value with standard deviation in each bin of magnetic latitude.  The green symbols are simulation predictions.
\label{fig:bkg_lat}}
\end{figure}

The background rate is a function of magnetic latitude on account of the flux of incident high energy charged particles.  In the orbital plane of PolarLight, high energy charged particles trapped in the Earth's magnetosphere result in high fluxes in the detector when it passes through the south Atlantic anomaly (SAA) or two polar regions (Figure~\ref{fig:map}). When the spacecraft approaches these regions, the detector is scheduled to power off for safety consideration.  Shown in Figure~\ref{fig:map} is the measured background rate in the energy range of 1--10 keV at different locations in the orbital plane on top of the flux map of trapped electrons.  It is obvious that the background rate increases if the spacecraft is close to regions of trapped electrons or at high latitudes. 

For observations executed at locations away from high flux regions, encircled by the dashed lines in Figure~\ref{fig:map}, we plot the background count rate as a function of magnetic latitude, shown in Figure~\ref{fig:bkg_lat}. The background count rate remains constant at $| \theta_{\rm m} | < 0.3$~rad and gradually increases with $| \theta_{\rm m} |$ otherwise. This indicates that the orbital variation of background is mild for PolarLight, within a factor of 1.5 from $\theta_{\rm m} = 0$ to $\pm0.6$~rad.  If we plot the background rate as a function of the magnetic parameter $L$ defined in \cite{Mcilwain1961}, our measurements span an $L$ range from 1 to 1.35 and the background flux has a power-law dependence of $2.51 \pm 0.13$ on $L$. 

As we co-added simulation results at different magnetic latitudes,  the simulated background flux fits the observation on average. However,  if we shift the simulation results at $| \theta_{\rm m} | < 0.3$~rad to match the data, the background flux is underestimated by a factor of $\sim$2 by simulations at $| \theta_{\rm m} | > 0.3$~rad. This is a direct consequence of the input model \citep{Mizuno2004}, in which the model flux for secondary electrons is higher at low latitudes.  The increase of background at high latitudes is consistent with measurements with other gaseous detectors \citep{Mason1983}.  Although the orbital variation of background flux cannot be perfectly reproduced, the variation trend in regions below and above  $| \theta_{\rm m} | \approx 0.3$, respectively, is the same between observations and simulations. 

\subsection{Background distribution on the detector plane}

\begin{figure}[t]
\centering
\includegraphics[width=0.8\columnwidth]{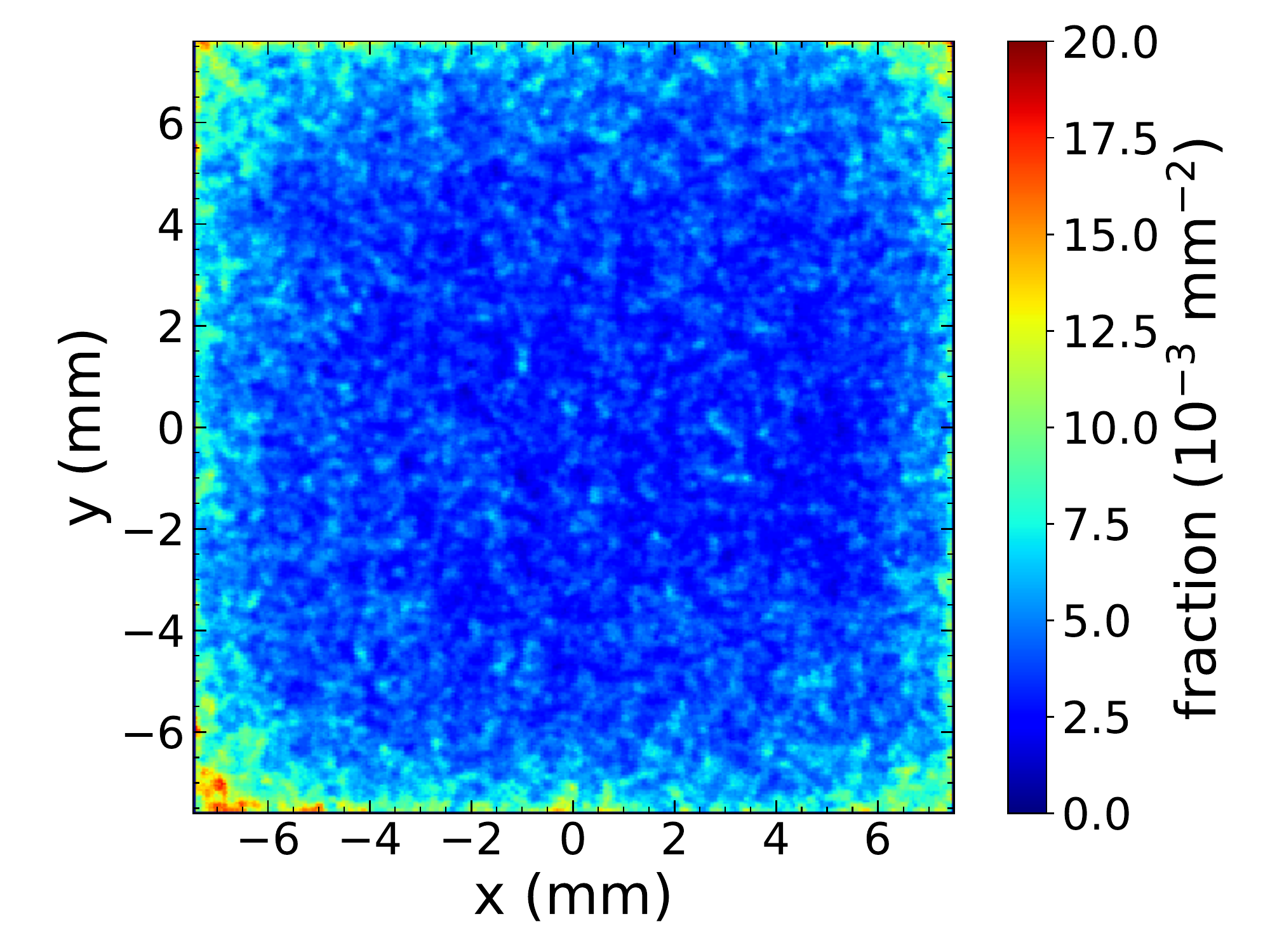}
\includegraphics[width=0.8\columnwidth]{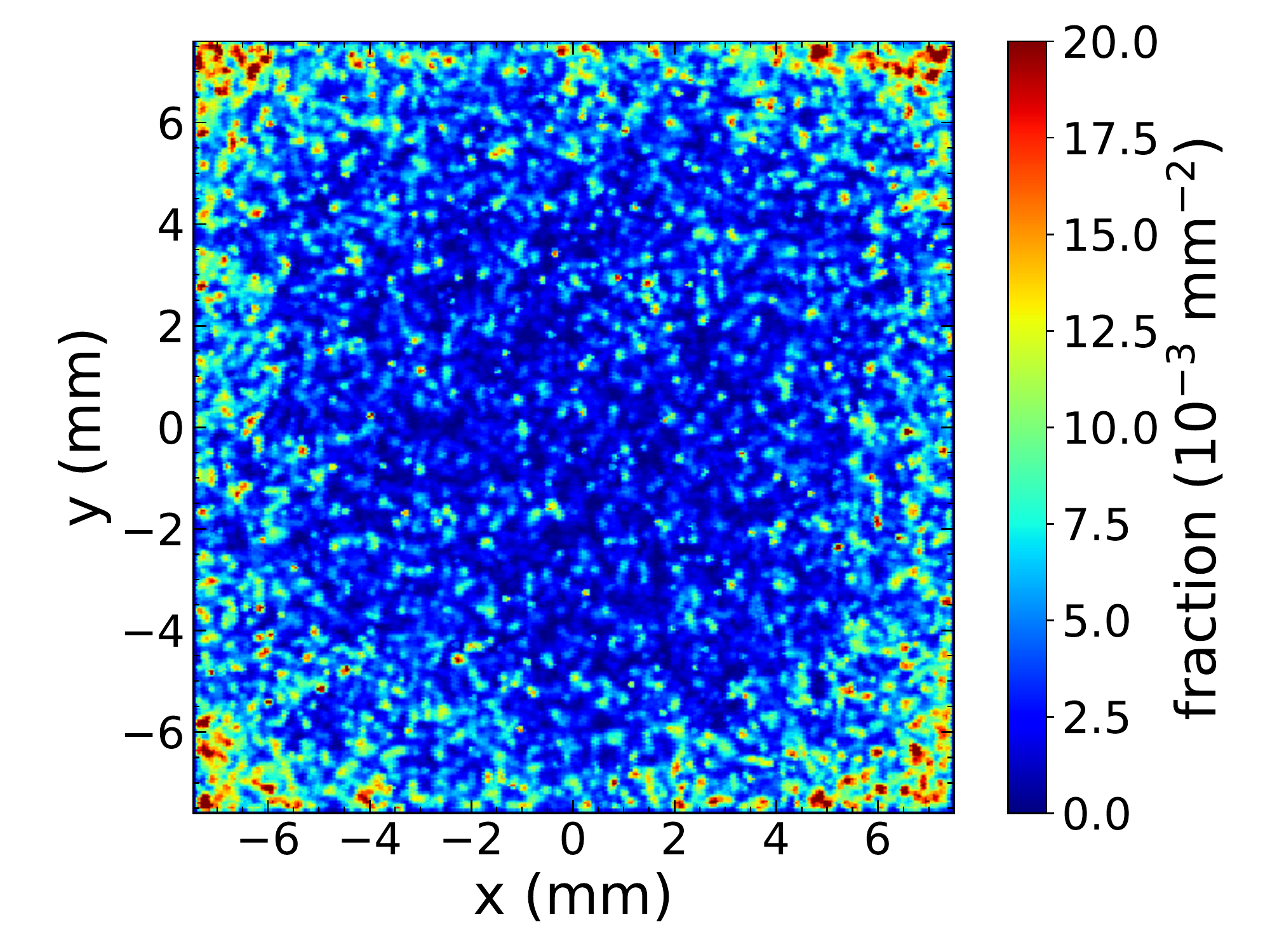}
\caption{Normalized hit maps for the measured background (\textbf{top}) and simulated background (\textbf{bottom}).  
\label{fig:hitmap}}
\end{figure}

\begin{figure}[t]
\centering
\includegraphics[width=0.8\columnwidth]{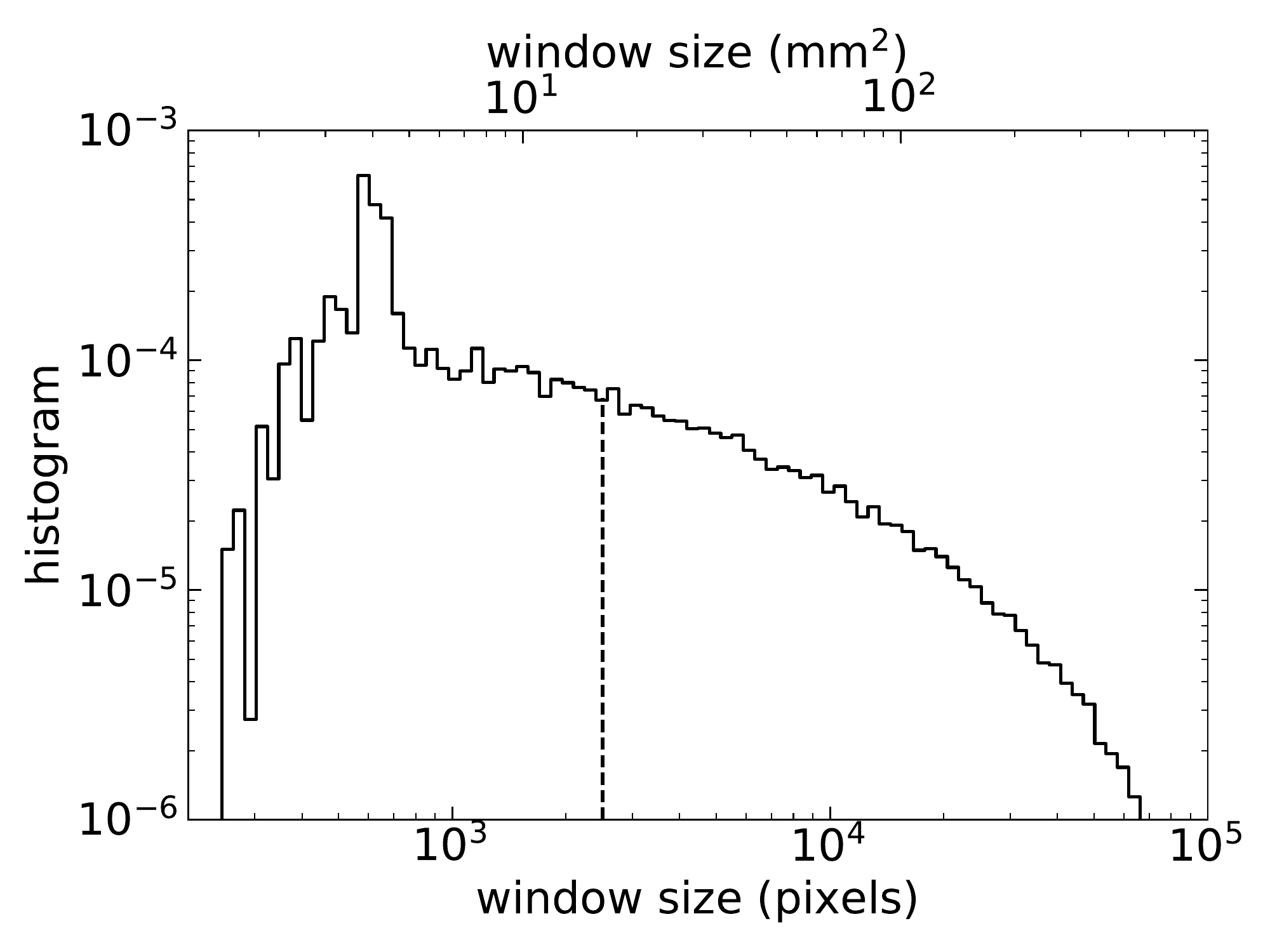}
\caption{Distribution of the event window size from simulations. The dashed line marks the threshold to separate valid events (used in Figure~\ref{fig:hitmap} bottom) and discarded events (used in Figure~\ref{fig:hitmap_cut}) . 
\label{fig:winsize}}
\end{figure}

\begin{figure}[t]
\centering
\includegraphics[width=0.8\columnwidth]{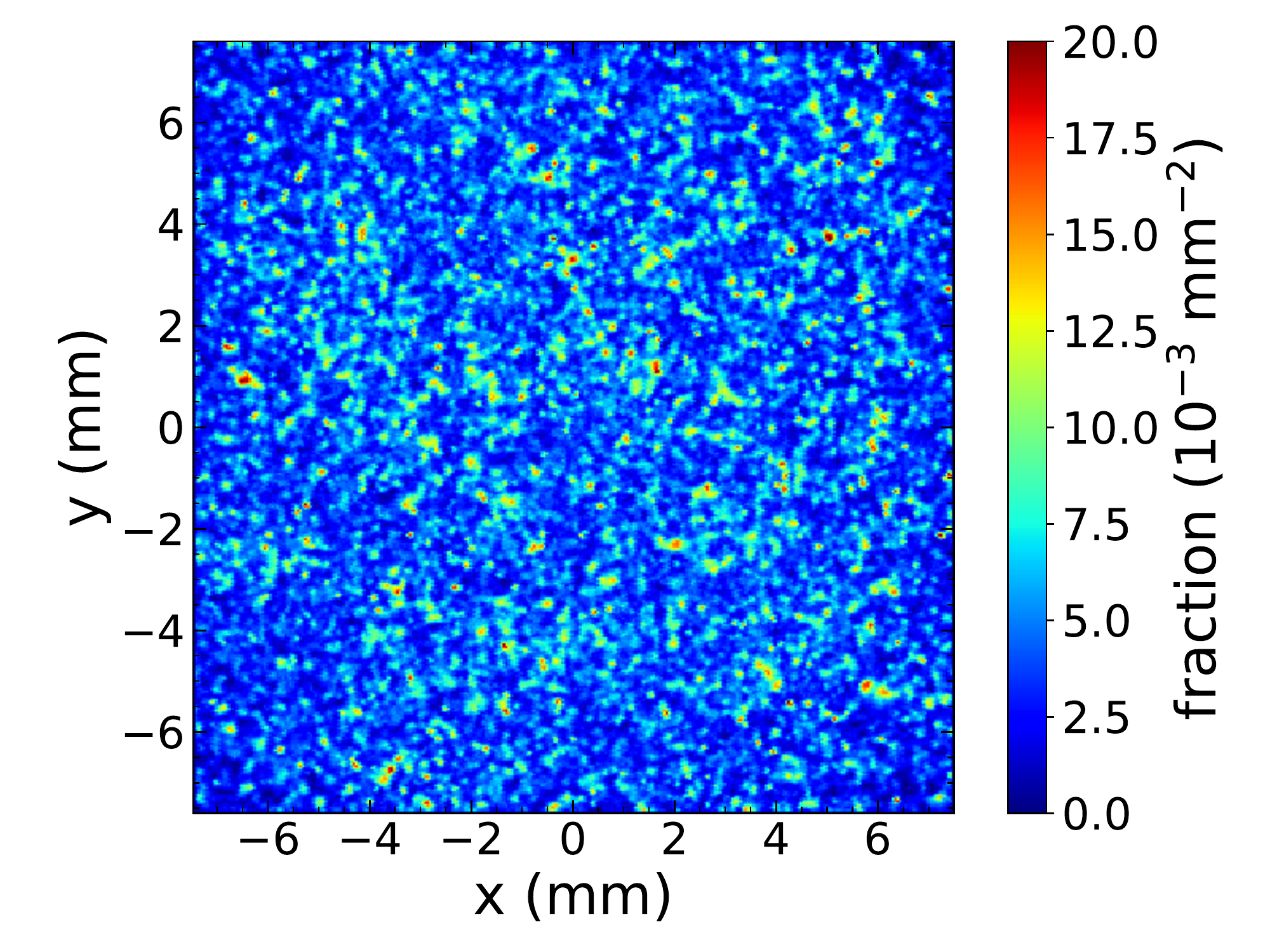}
\caption{Normalized hit map constructed using simulated events that are discarded due to large window sizes. 
\label{fig:hitmap_cut}}
\end{figure}

Background events are not evenly distributed on the detector plane. Figure~\ref{fig:hitmap} shows images of energy deposits on each pixel (hereafter called the hit map) due to background events in the energy range of 1--10 keV on the detector plane from the measured and simulated data, respectively.  The energy deposit, or equivalently the event rate, is enhanced at the edges and corners of the ASIC chip, consistent with laboratory results \citep{Soffitta2012}.   Examination of the simulation reveals that such an edge effect is not due to enhancement of secondary particles near the edges.  Rather, it is because of the trigger and readout algorithm.  For each event, a rectangle readout window is defined to contain all of the triggered pixels plus some margins on each side.  The readout electronics will discard an event if its window size (number of pixels in the window) is too large.  The distribution of window size with the threshold for valid events is shown in Figure~\ref{fig:winsize}. Therefore, background events due to high energy charged particles, which tend to produce long tracks, are more likely to be recorded if they hit the gas above the edge or corner of the ASIC with a partial energy deposit and correspondingly a relatively short track and small window.  The hit map for discarded events, those with large windows, is shown in Figure~\ref{fig:hitmap_cut} for comparison and no longer displays an enhancement near the edge.  

\begin{figure}[t]
\centering
\includegraphics[width=0.8\columnwidth]{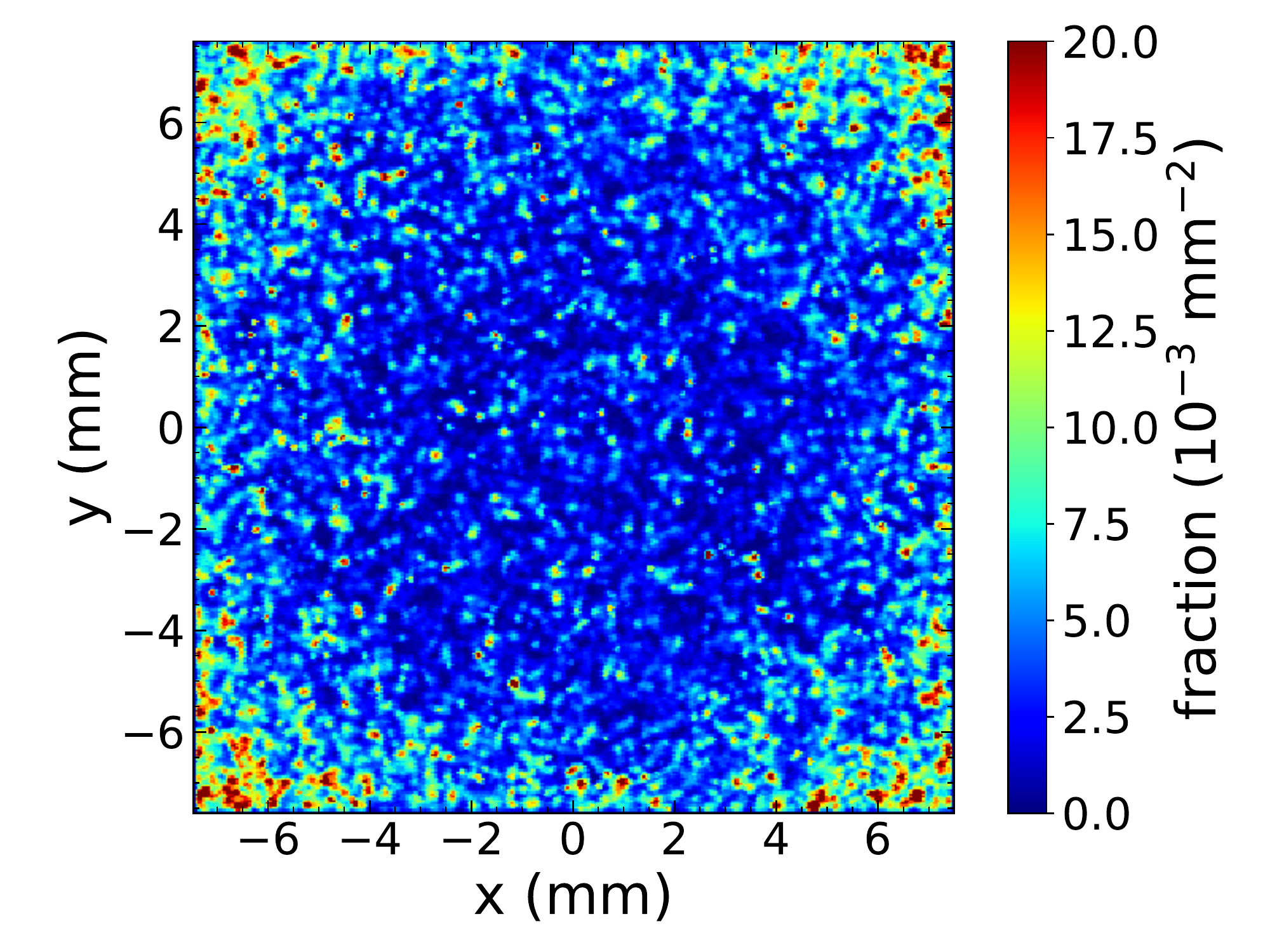}
\caption{Normalized hit map from simulations with a larger chamber, which is 70~mm $\times$ 70~mm, vs.\ 27~mm $\times$ 27~mm for PolarLight.  
\label{fig:hitmap_large}}
\end{figure}

To verify this, we also make simulations assuming a larger chamber  (70~mm $\times$ 70~mm vs.\ 27~mm $\times$ 27~mm in the current setup) , in which case the chamber materials are farther away from the sensitive volume.  The hit map is shown in Figure~\ref{fig:hitmap_large}. As one can see, the background distribution is almost identical to that with the smaller chamber.  

Since events near the edge may have partial energy deposits, their energy spectrum and angular modulation are distorted. In practice, events near the edges are not used in scientific analysis for PolarLight.

\section{Discussions and conclusions}
\label{sec:discussion}

In this work, we have constructed a mass model for the CubeSat that contains PolarLight and performed particle tracking simulations to reproduce the in-orbit background.  The simulated energy spectra and event images suggest that high energy electrons in the orbit are the dominant source of background, while protons plus the CXB contributes roughly 19\% to the background counts in the energy range of 1-10 keV. This conclusion is further confirmed by comparing the simulated event images with measurements. However, it is contradictory to the \citet{Mizuno2004} model, which predicts that the ratio of background counts induced by protons to that induced by electrons is 1.7 in 1--10 keV, while the best-fit model decomposition suggests a ratio of 0.18 instead. The in-orbit background varies with orbital locations, being minimum at magnetic latitudes within $\pm 0.3$~rad and enhanced by a factor of $\sim$1.5 at magnetic latitudes of $\pm$0.6~rad.  Again, it is inconsistent with model predictions (see Figure~\ref{fig:bkg_lat}) but in line  with measurements with other instruments.   It is unclear if these deviations are a result of the different orbits.

On the detector plane, the background distribution is uneven, with more events detected near the edges. This is not because there are more background events toward the edges, rather, it is because background events, or those with long tracks, have a lower probability of being recorded  around the center due to a rule in the readout electronics.  When we check the simulation results in the whole chamber, we find that the 1-10 keV background density near the corner of the chamber (not the corner of the ASIC) is slightly higher than the central value by $\sim$20\% or so.  Above the ASIC region, the background variation is small enough, further suggesting that the readout algorithm is the dominant cause of the edge effect seen in Figure~\ref{fig:hitmap}.

Four different morphology types are defined for the background events. Compared with images produced by X-rays, two of them (type (b) and (c)) show distinct features while the other two (type (a) and (d)) are due to identical physical processes, i.e., ionized by electrons of several keV.  Events of types (a) and (d) occupy a fraction of roughly 30\% in 1--10 keV or 28\% in 2--8 keV, and cannot be discriminated from X-ray events by any means.  The total background rate in the energy band of 2--8 keV is measured to be about $2 \times 10^{-2}$~\ctscm\ in the central region. With an effective particle discrimination algorithm (removing 72\% of the background), the residual background can be reduced to about $6 \times 10^{-3}$~\ctscm. For future imaging X-ray polarimeters like XIPE or eXTP, the focal spot size is on the order of mm. Thus, the internal background should be negligible. However, there are more surrounding masses for those large missions, and may cause a higher, but still insignificant, internal background.   

As the background is not a strong function of latitude, a very low inclination orbit can help improve the observing efficiency by reducing SAA passages, but will not help depress the background.   The magnetic dependence of the background rate is somewhat steeper than the result obtained using other gaseous detectors \citep{Mason1983}, a power-law index of about 1.5 between $L = 1 - 3$ vs.\ 2.5 in our case.  This is perhaps because our chamber is rather small and events with long tracks are automatically discarded by the readout electronics, or simply due to a narrow $L$ range in our measurements.

This work validates the mass model of PolarLight. Due to a low background rate, the measured background does not have a sufficiently high statistic allowing for a meaningful analysis of their polarimetric modulation. This can be done in the future when we have accumulated enough data for the background. Developing an effective algorithm for particle discrimination is also of essential importance, and will be a topic to study following this work.  

\acknowledgements We thank the anonymous referee for useful comments. HF acknowledges funding support from the National Natural Science Foundation of China under the grant Nos.\ 11633003, 12025301, and 11821303, the CAS Strategic Priority Program on Space Science (grant No.\ XDA15020501-02), and the National Key R\&D Project (grants Nos.\ 2016YFA040080X \& 2018YFA0404502).  


\end{document}